\RequirePackage[2020-02-02]{latexrelease} 

\documentclass[a4paper,fleqn,usenatbib]{mnras}
\usepackage{longtable}

\usepackage[switch]{lineno}
\usepackage{natbib}
\bibpunct{(}{)}{;}{a}{,}{,}

\usepackage{xspace}
\usepackage{pstricks}
\usepackage{color}
\usepackage{pgf}
\graphicspath{{./}}

\usepackage{hyperref}
\hypersetup{
    unicode=true,           
    pdftoolbar=true,        
    pdfmenubar=true,        
    pdffitwindow=true,      
    pdftitle={Speckle observations of Kalliope},
    pdfauthor={},
    pdfsubject={Planetary Science},
    pdfkeywords={},         
    pdfnewwindow=true,      
    colorlinks=true,        
    linkcolor=gray,         
    citecolor=blue,         
    filecolor=gray,         
    urlcolor=gray           
} 

\newcommand{\genoid}{\texttt{Genoid}\xspace}
\newcommand{\xitau}{\texttt{xitau}\xspace}
\newcommand{\miriade}{\texttt{Miriade}\xspace}
\newcommand{\ssodnet}{\texttt{SsODNet}\xspace}
\newcommand{\astropy}{\texttt{astropy}\xspace}

\usepackage{graphicx}
\usepackage{amsmath,amssymb}
\newcommand{\be}{\begin{equation}}
\newcommand{\ee}{\end{equation}}

\usepackage{catoptions}
\makeatletter

\def\Autoref#1{%
  \begingroup
  \edef\reserved@a{\cpttrimspaces{#1}}%
  \ifcsndefTF{r@#1}{%
    \xaftercsname{\expandafter\testreftype\@fourthoffive}
      {r@\reserved@a}.\\{#1}%
  }{%
    \ref{#1}%
  }%
  \endgroup
}
\def\testreftype#1.#2\\#3{%
  \ifcsndefTF{#1autorefname}{%
    \def\reserved@a##1##2\@nil{%
      \uppercase{\def\ref@name{##1}}%
      \csn@edef{#1autorefname}{\ref@name##2}%
      \autoref{#3}%
    }%
    \reserved@a#1\@nil
  }{%
    \autoref{#3}%
  }%
}
\makeatother

\title{Speckle observations of the binary asteroid (22) Kalliope with C2PU/PISCO}

\author[E. Aristidi et al.]{E. Aristidi,$^{1}$\thanks{E-mail:eric.aristidi@oca.eu},
B. Carry$^{1}$,
K. Minker$^{1}$,
J.-L. Prieur$^{2,3}$,
M. Scardia$^{4,5}$, 
J.-P. Rivet$^{1}$,
\newauthor
P. Bendjoya$^{1}$
L. Abe$^{1}$, 
R.-W. Argyle$^{6}$, 
L. Koechlin$^{2,3}$, 
J.F. Ling$^{7}$, 
L. Maccarini$^{8}$, 
\newauthor
L. Pansecchi$^{5}$, 
L. Piccotti$^{7,9}$, 
J. S\'erot$^{10}$,
D. Vernet$^{4}$\\ 
$^{1}$Universit\'e C\^ote d'Azur, Observatoire de la C\^ote d'Azur, CNRS, Laboratoire Lagrange, France\\
$^{2}$Université de Toulouse – UPS-OMP – IRAP, Toulouse, France\\
$^{3}$CNRS – IRAP, 14 avenue Edouard Belin, 31400 Toulouse, France\\
$^{4}$Université C\^ote d’Azur, Observatoire de la Côte d’Azur - C2PU, Nice, France\\
$^{5}$INAF – Osservatorio Astronomico di Brera, Via E. Bianchi 46, 23807 Merate, Italy\\
$^{6}$Institute of Astronomy, Madingley Road, Cambridge, CB3 0HA, United Kingdom\\
$^{7}$Observatorio Astronomico R.M. Aller, P.O. Box 197, Universidad de Santiago de Compostela, Spain\\
$^{8}$Desio, Monza and Brianza, Italy\\
$^{9}$Instituto Universitario de Investigaci\'on en Matematicas y Aplicaciones, Universidad de Zaragoza, C. de Pedro Cerbuna,\\ 12, 50009, Zaragoza, Spain\\
$^{10}$Institut Pascal – CNRS UMR 6602, Avenue B. Pascal 63178 Aubi\`ere, France\\
}
\date{Accepted XXX. Received YYY; in original form ZZZ}
\pubyear{2023}

\begin{document}
\label{firstpage}
\pagerange{\pageref{firstpage}--\pageref{lastpage}}
\maketitle

\begin{abstract}
We present new speckle measurements of the position of Linus, the satellite of the asteroid (22) Kalliope, obtained at the 1m C2PU-Epsilon telescope on the  {\textit{Plateau de Calern}}, France. Observations were made in the visible domain with the speckle camera PISCO. We obtained 122 measurements in February--March 2022 and April 2023, with a mean uncertainty close to 10 milli-arcseconds on the angular separation.

\end{abstract}



\section{Introduction}
\label{par:intro}
The main-belt asteroid (22) Kalliope is classified as M-type, thought to be linked with iron meteorites \citep{1989aste.conf.1139T, 2022A&A...665A..26M}. It has a small companion Linus, which was discovered in 2001 \citep{Merline2001, Margot2001} by means of adaptive optics (AO) on large ground-based telescopes (3.6\,m CFHT and 10\,m Keck). Since 2001, astrometric measurements of Linus have been measured by several authors using different techniques (AO imaging, stellar occultations, speckle interferometry), leading to a set of 188 measurements \citep[see][and references therein for more details]{Ferrais2022}. The orbit of Linus has been studied by various authors and methods as the number and quality of available measurements increased \citep[see][]{2008Icar..196..578D, 2008Icar..196...97M, Vachier2012}. The latest analysis, on the largest set of observations, was published by 
\citet{Ferrais2022}, using two different orbit-fitting algorithms, \genoid \citep{Vachier2012} and \xitau \citep{2021A&A...653A..56B}. 
Although the shape of (22) Kalliope is well modeled and highly non-spherical, the orbit found by these authors is compatible with a purely Keplerian motion, suggesting that Kalliope could have a differentiated interior \citep{Ferrais2022}. The orbit is also compatible with a multipole model of the shape of (22) Kalliope, suggesting a homogeneous composition \citep{Ferrais2022}. Since the internal structure of (22) Kalliope is currently unclear, monitoring the orbit of Linus is of high interest because the long-term dynamics of the system can reveal small perturbations due to the internal structure of Kalliope. 


In { this} study, we present new speckle observations of Kalliope with the Epsilon telescope (diameter 104\,cm) of the
{\textit{Centre Pédagogique Planètes et Univers}}
(C2PU) facility \citep{Bendjoya12}
located at the Calern observing site (Côte d'Azur Observatory, France, IAU code: 010). This telescope is equipped with the Pupil Interferometry Speckle camera and COronagraph (PISCO) instrument which provides high angular resolution observations in the visible domain. The instrumentation is described in \cite{Scardia2019}. Over the last 30 years, PISCO has produced thousands of measurements of binary star astrometry at the milli-arcsec precision level. Its smallest accessible angular separation is about 130~milli-arcsec (mas) and the limiting magnitude is $V\simeq 15$ (for the dimmer component of a binary) in good seeing conditions (i.e. seeing value below 1~arcsec).

Observations were performed between December 2021 and April 2023. This period corresponded to exceptional conditions for Kalliope visibility at Calern, since it was in opposition and at maximum declination ($\delta\simeq +34^\circ$ at the beginning of 2022). The pair Kalliope-Linus is a difficult target for speckle on a 1.04~meter telescope owing to its apparent magnitude ($V > 11$) and the large magnitude difference with its satellite ($\Delta V\simeq 4$). It is often beyond the limits of our instrumentation, but during the period 2021--2023, its visual magnitude $V\simeq 11$ and expected angular separation (100~mas $< \rho <$ 700~mas) made it accessible to PISCO and we attempted observations.

Four observing runs were conducted during December 30-31, 2021; February 10-20, 2022;  March 21-25, 2022 and April 5-19, 2023. The first run gave 19 astrometric measurements of Linus position respective to Kalliope which were published in \cite{Ferrais2022}, they are indeed the first successful observations of the pair Kalliope-Linus with PISCO. This makes C2PU the smallest telescope that directly measured the astrometry of a satellite of an asteroid.
We report here on the three other observing runs and the astrometry of Linus.

\begin{figure*}
\input{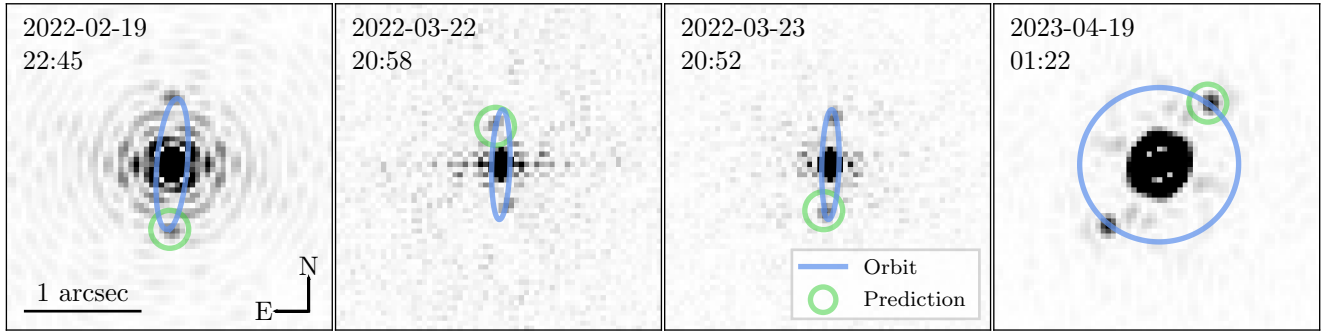}
\caption{Exemples of 2D autocorrelations of Kalliope speckle images. The orbit of Linus and predicted position at the time of observation from
\citet{Ferrais2022} are overplotted. Note the { difference in the position angle between the two middle graphs, one day apart:  the period of Linus is approximately 3.6 days, which imply a motion of $\sim$100 degree per day. Taking into account the inclination of the orbit, we observe a difference of almost 180$^\circ$ in the position angle.}}
\label{fig:ac}
\end{figure*}

\begin{figure*}
\input{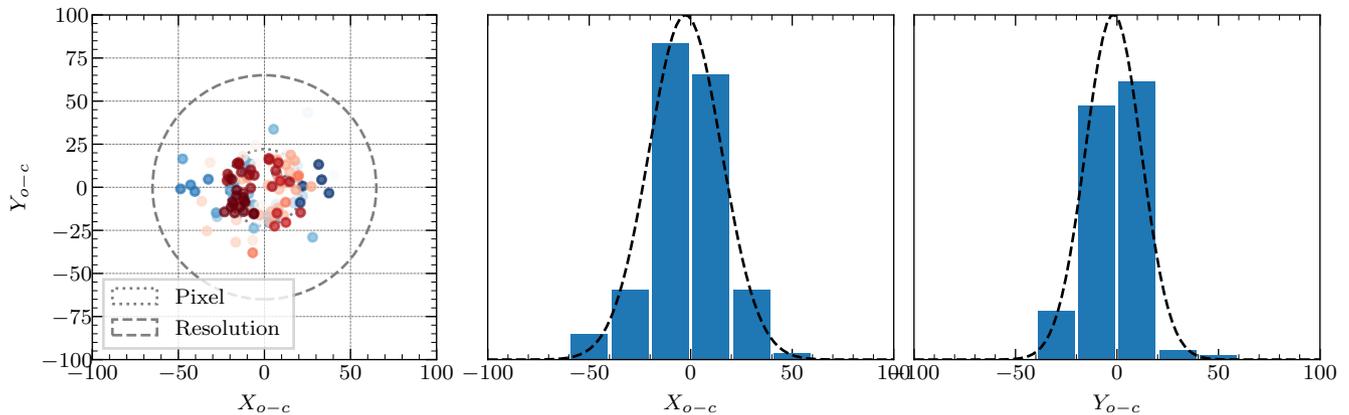}
  \caption{Residuals of the position measurements relative to the published orbit (O-C). \textbf{Left:} scatter plot in the $XY$ plane (blue points from 2022, red from 2023). We indicate the pixel size of 2022 (55\,mas) and angular resolution of the telescope (130\,mas). \textbf{Center:} histogram of the residuals in the $X$ direction (right ascension). The residuals follow a Gaussian distribution centered on zero. \textbf{Right:} same for the $Y$ direction (declination). 
  \label{fig:OmC}}
\end{figure*}

\begin{figure*}
\input{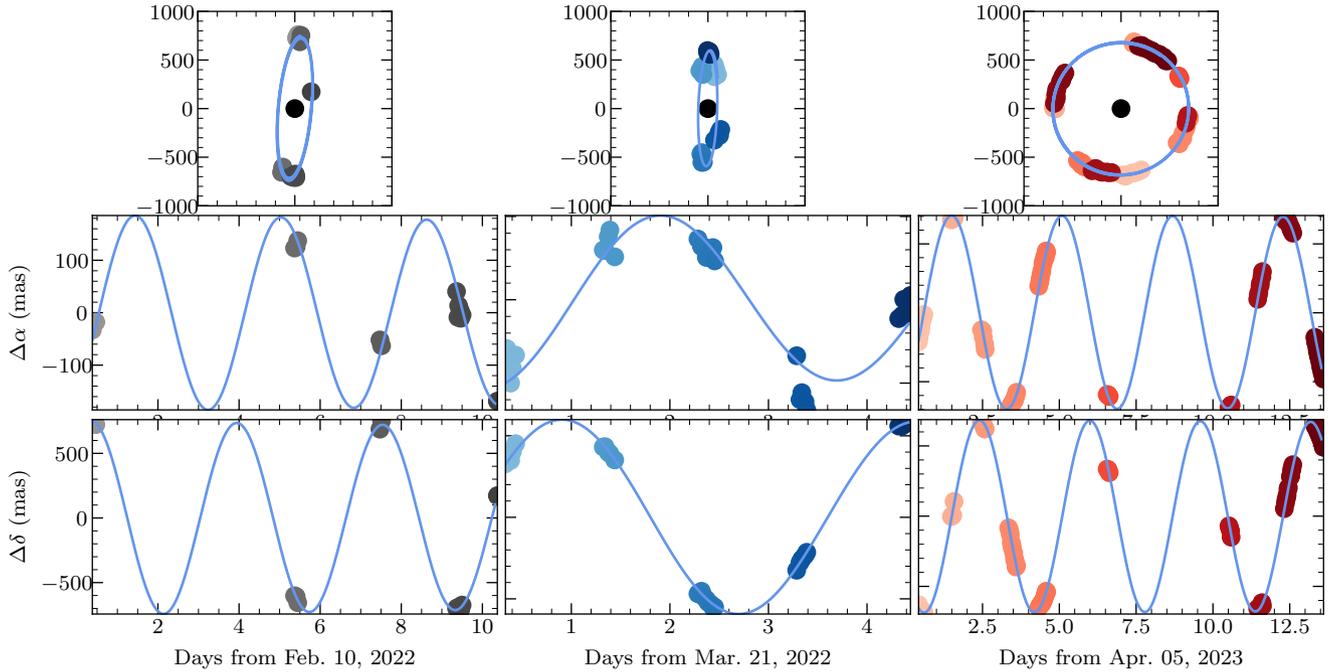}
\caption{Comparison of the measured positions of Linus with respect to Kalliope
  ($\Delta \alpha$, $\Delta \delta$, same color scale as Fig.~2) 
  and its predicted orbit in blue, 
  for February 2022 (left), March 2022 (center), and April 2023 (right) data sets.
The first row presents the positions on 
the plane of the sky (North is up, East is right). Linus orbits counter-clockwise in 2022 and clockwise in 2023.}\label{fig:orbit} 
\end{figure*}

\section{Observations and data processing}
\label{par:obs}
Observations were made with the PISCO instrument equipped with { a} back-illuminated EMCCD detector from ANDOR (iXon Ultra 897 in 2022, iXon Ultra 888 in 2023).
It provides cubes of $\sim 15\, 000$ images taken with a short exposure time to freeze atmospheric turbulence. Images were recorded without filter in the whole visible domain (central wavelength $\lambda = 550$\,nm, bandpass $\Delta\lambda= 500$\,nm) with an exposure time of 100 ms. This value of the exposure time is very long for speckle data; it is generally of the order of magnitude of the coherence time, typically 5–20\,ms, 
but it was selected as a trade-off between the low signal-to-noise ratio (SNR) of short exposure frames and the loss of information caused by the turbulence in long exposure images. The spatial sampling was 54.8\,$\pm$\,0.4 milli-arcseconds/pixel in 2022, and 44.3\,$\pm$\,0.3 milli-arcseconds/pixel for the 2023 run.

For each measurement, a nearby reference star of same magnitude as Kalliope (or slightly brighter) was observed. Data were processed via Labeyrie's speckle interferometry technique \citep{Labeyrie70}. This results in 2D visibilities and auto-correlation (AC) functions, as shown in \Autoref{fig:ac}. The AC shows a central bright peak and two smaller ones. The position of these secondary peaks gives the position vector of the companion, they are measured with an interactive program that fitted and subtracted the residual background \citep{Scardia2007}. There is, however, a 180$^\circ$ ambiguity since the two secondary peaks are symmetric. Solving this ambiguity requires the use of alternative speckle techniques \citep{Aristidi97, Bagnuolo88, KnoxThomson74} which were not efficient here because of the low SNR of visibilities. Instead, we compared the observed positions with the predicted one and chose the closest one.

We made a total of 151 observations of the system, among which 122 were successful and provided measurements of the astrometry of Linus. The position of Kalliope in the sky was optimal (airmass lower than 1.5 most of the time), but we faced episodes of strong wind and bad seeing (larger than 2~arcsec) which made the detection of Linus difficult or sometimes impossible. Also in February 2022 (and the beginning of April 2023) we were close to the full Moon and the high sky background affected the observation of the companion.

The astrometry of Linus relative to Kalliope in February--March 2022 and April 2023 is presented in \Autoref{table:mes}. 
The median formal uncertainty on the determination of the position of Linus is of the order of 10\,mas ($\simeq$ 0.2~pixel). This uncertainty comes
from the determination of the centres of the autocorrelation peaks \citep{Scardia2007}. Note that the disc of Kalliope was unresolved at the time of the observations (angular diameter $\le$100~mas).

We present in \Autoref{fig:OmC} the residuals of the measurements with the predictions
from the Keplerian orbit presented by \citet{Ferrais2022}. The { arithmetic} mean values are $\langle X_{O-C}\rangle=-2.6$\,mas and $\langle Y_{O-C}\rangle=-1.8$\,mas,
with $X$ and $Y$ representing the { differential  coordinate} ``Linus minus Kalliope'' along
right ascension and declination, respectively, and the indices $O$ and $C$ standing
for observed and computed positions.
These values are 
much smaller than the measurement uncertainty, indicative of a very good agreement between the orbital solution of \citet{Ferrais2022} and these new observations. The dispersion (standard deviation) of these residuals are $\sigma_{X,O-C}=17.9$\,mas and $\sigma_{Y,O-C}=13.8$\,mas. These values may appear quite large compared with the RMS residuals published in the Table~2 of~\citet{Ferrais2022}. However, most measurements listed in that paper were made using 8\,m class telescopes. The residuals presented here remain, however, much smaller than the angular resolution of a 1\,m telescope in the visible (130\,mas) and the pixel scale ($\sim$45-50\,mas/pixel). 


\Autoref{fig:orbit} shows the measured positions $X$ and $Y$ together with the arc of orbit predicted by the ephemeris, for the runs of 2022 and 2023. In 2022 the orbit was seen under a small angle from its plane and the pair was not resolved when Linus was close to the periapsis (angular separation $\simeq$ 100mas).
Therefore most of the 2022 measurements were made where Linus was close to its maximum angular separation. In 2023, the orbit was seen almost pole on and the astrometry was hence easier to measure, the angular separation being constant. The orbit fits very well the data points in the $Y$ direction (declination). The agreement in the $X$ direction (right ascension) is also satisfactory, though a few discrepancies were observed for the March 2022 observing run (this is why the standard deviation of the residuals is larger in $X$ than in $Y$). They might be produced by artifacts in the speckle AC functions: the amplitude of the secondary peaks of the AC is very faint and can be contaminated by noise. It is also possible that the orbital solution needs refinement, but more observations are needed in the future.
Favorable conditions for Kalliope observations in the Northern hemisphere, with high declination, will be meet again in 2026, 2031, and almost every 5 years in the future.

\section{Conclusion}
\label{par:concl}
We present new astrometric observations of Linus, the satellite of (22) Kalliope, by means of speckle interferometry using the 1.04~meter C2PU telescope. Our data set or 122 measurements complements the 188 published positions of Linus \citep{Ferrais2022}, and adds 16 months to the temporal coverage. The results are consistent with the published orbital solution, with mean residuals smaller than 3\,mas. 

Observing a satellite of an asteroid with a 1\,m telescope is quite a challenge. It is nevertheless feasible, which encourages us to attempt more observations in the future. The main advantage of using a small telescope like C2PU is the possibility to observe during long periods of time and perform a continuous monitoring of the astrometry of binary asteroids.
In the case of (22) Kalliope, this opens the possibility of long-term study of its dynamics to constrain further its internal structure.

\section*{Acknowledgements}
The authors wish to thank P. Tanga from Lagrange {Laboratory, \textit{Observatoire de la C{\^o}te d'Azur}} for allowing us to use his Andor iXon Ultra 897 camera with PISCO. Thanks also to F. Vachier from the IMCCE, (\textit{Observatoire de Paris}) for useful discussions about the orbital solution, and to D. Bonneau (\textit{Observatoire de la C{\^o}te d'Azur}) for suggestions about the manuscript.
This research used the
\miriade\footnote{\url{https://ssp.imcce.fr/webservices/miriade/}}
\citep{2009-EPSC-Berthier}
and
\ssodnet\footnote{\url{https://ssp.imcce.fr/webservices/ssodnet}} 
\citep{2023A&A...671A.151B}
Virtual Observatory tools,
and
\astropy\footnote{\url{http://www.astropy.org}},
a community-developed core Python package and an ecosystem of tools 
and resources for astronomy \citep{astropy:2013, astropy:2018, astropy:2022}. 
  Thanks to all the developers and maintainers.

\section*{Data availability statement}
The data underlying this article are available in the article (Table 1).
\bibliographystyle{mnras}
\bibliography{biblio}

\renewcommand{\arraystretch}{0.7} 
{\onecolumn 
\small
\begin{longtable}{cccccccccc}
\caption{Astrometry of Linus with respect to Kalliope along Right ascension ($X$) and Declination ($Y$), together with the formal uncertainty ($\sigma$) and residuals ($O-C$) with the orbit by \citet{Ferrais2022}. {\bf The column Vmag is the visual magnitude predicted by the ephemeris.}} \label{table:mes}\\
\hline 
Date & UTC & MJD & Airmass & Vmag & $X_O$  & $Y_O$  & $\sigma$ &  $X_{O-C}$ & $Y_{O-C}$ \\  
    &     &          &   &   & [mas]  & [mas]  & [mas]    & [mas]      & [mas] \\ \hline
\endfirsthead
\multicolumn{9}{c}%
{\tablename\ \thetable\ -- \textit{Continued from previous page}} \\
\hline
Date & UTC & MJD & Airmass & Vmag & $X_O$  & $Y_O$  & $\sigma$ &  $X_{O-C}$ & $Y_{O-C}$ \\  
    &     &          &   &   & [mas]  & [mas]  & [mas]    & [mas]      & [mas]     \\ \hline
\endhead
\hline \multicolumn{9}{r}{\textit{Continued on next page}} \\
\endfoot
\hline
\endlastfoot

2022-02-10 & 21:10:55.3   & 59620.882581 &1.0 & 11.1 & -33 & 761  &  11 & 19  &  12    \\
2022-02-10 & 22:15:57.6   & 59620.927743 &1.1 & 11.1 & -20 & 734  &  11 & 17  & -8     \\
2022-02-10 & 23:15:00.2   & 59620.968750 &1.2 & 11.1 & -18 & 723  &  11 &  7  & -10    \\
2022-02-15 & 20:51:08.1   & 59625.868843 &1.0 &  11.2& 123 & -602 &  11 & -28 & -17    \\
2022-02-15 & 21:47:54.8   & 59625.908264 &1.1 & 11.2 & 125 & -608 &  11 & -18 &   5    \\
2022-02-15 & 22:43:47.4   & 59625.947072 &1.2 & 11.2 & 137 & -655 &  11 & 2   & -17    \\
2022-02-17 & 23:22:24.5   & 59627.973889 &1.3 &11.2  & -52 & 687  &  16 & 28  & -29    \\
2022-02-18 & 00:19:51.2   & 59628.013785 &1.5 &11.2  & -63 & 754  &  11 &  5  &  33    \\
2022-02-19 & 20:41:53.2   & 59629.862419 &1.0 & 11.3 & 40  & -699 &  11 & -9  &  13    \\
2022-02-19 & 21:30:24.7   & 59629.896111 &1.1 & 11.3 & -9  & -693 &  16 & -48 &  16    \\
2022-02-19 & 21:56:13.5   & 59629.914039 &1.1 & 11.3 & 13  & -708 &  11 & -20 &   2    \\
2022-02-19 & 22:46:17.9   & 59629.948808 &1.2 & 11.3 & 5   & -695 &  11 & -18 &   3    \\
2022-02-19 & 23:11:01.8   & 59629.965984 &1.3 & 11.3 & -11 & -709 &  11 & -28 & -15    \\
2022-02-19 & 23:59:16.3   & 59629.999491 &1.5 & 11.3 & -4  & -674 &  11 & -11 &   8    \\
2022-02-20 & 20:51:14.8   & 59630.868912 &1.5 & 11.3 & -169& 173  &   8 &   7 &   8    \\
2022-03-21 & 19:57:13.0   & 59659.831400 &1.1 & 11.8 & -67 & 328  &  16 &  33 & 4      \\
2022-03-21 & 20:22:42.6   & 59659.849097 &1.1 & 11.8 & -59 & 344  &  27 &  40 &  7     \\
2022-03-21 & 21:13:01.7   & 59659.884039 &1.3 & 11.8 &-100 & 349  &  16 &  -4 & -20    \\
2022-03-21 & 21:38:29.6   & 59659.901725 &1.3 & 11.8 & -82 & 402  &  9 &  12 & 19   \\
2022-03-21 & 22:25:54.1   & 59659.934653 &1.6 & 11.8 & -66 & 452  &  11 &  25 & 42   \\
2022-03-22 & 19:46:25.9   & 59660.823900 &1.1 & 11.8 & 60  & 431  &  11 &   8 & -14  \\
2022-03-22 & 20:11:49.1   & 59660.841539 &1.1 & 11.8 & 59  & 433  &  11 &   5 & 0    \\
2022-03-22 & 20:58:19.4   & 59660.873831 &1.2 & 11.8 & 76  & 400  &  11 &  16 & -9   \\
2022-03-22 & 21:23:03.2   & 59660.891007 &1.3 & 11.8 & 84  & 390  &  16 &  22 & 5    \\
2022-03-22 & 22:33:43.2   & 59660.940081 &1.6 & 11.8 & 52  & 352  &  11 & -17 & -4   \\
2022-03-23 & 18:49:51.3   & 59661.784618 &1.1 & 11.8 & 73  & -469 &  11 & -6  & -24  \\
2022-03-23 & 19:41:22.6   & 59661.820394 &1.1 & 11.8 & 64  & -454 &  11 & -11 & 15   \\
2022-03-23 & 20:06:05.1   & 59661.837558 &1.1 & 11.8 & 64  & -484 &  11 & -9  & 4    \\
2022-03-23 & 20:52:12.7   & 59661.869583 &1.2 & 11.8 & 51  & -503 &  11 & -18 & -5   \\
2022-03-23 & 21:16:53.1   & 59661.886725 &1.3 & 11.8 & 55  & -521 &  11  & -12 & -13 \\
2022-03-23 & 22:03:17.2   & 59661.918947 &1.5 & 11.8 & 56  & -535 &  11 &  -7 & -10  \\
2022-03-23 & 22:28:14.6   & 59661.936273 &1.6 & 11.8 & 63  & -553 &  11 &   3 & -21  \\
2022-03-23 & 22:53:29.5   & 59661.953808 &1.8 & 11.8 & 47  & -551 &  11  & -11 & -11 \\
2022-03-24 & 18:52:18.1   & 59662.786319 &1.0 & 11.8 & -67 & -325 &  11 &   6 & -17  \\
2022-03-24 & 19:40:29.9   & 59662.819780 &1.1 & 11.8 & -120& -277 &  16 & -43 & 1    \\
2022-03-24 & 20:07:16.4   & 59662.838380 &1.1 & 11.8 & -112& -256 &  16 & -33 & 4    \\
2022-03-24 & 20:54:07.9   & 59662.870914 &1.2 & 11.8 & -122& -233 &  16 & -40 & -3   \\
2022-03-24 & 21:19:19.4   & 59662.888414 &1.3 & 11.8 & -132& -215 &  16 & -49 & -1   \\
2022-03-25 & 19:43:23.7   & 59663.821794 &1.2 & 11.8 & -23 & 560  &  11 & 22  & 0    \\
2022-03-25 & 20:31:46.7   & 59663.855394 &1.2 & 11.8 & -19 & 560  &  11 &  20 & -9   \\
2022-03-25 & 20:56:58.2   & 59663.872894 &1.3 & 11.8 & 0   &  569 &  11 &  37 & -4   \\
2022-03-25 & 21:43:43.3   & 59663.905359 &1.4 & 11.8 & 1   &  584 &  11 &  33 & 4    \\
2022-03-25 & 22:09:00.2   & 59663.922917 &1.5 & 11.8 & -16 & 586  &  16 &  12 & 4    \\
2022-03-25 & 22:34:25.2   & 59663.940567 &1.7 & 11.8 & 5   &  597 &  11 &  31 & 13   \\ 
2023-04-05 & 22:01:45.8   & 60039.917882 &1.3 & 11.1 & -210  &  -630 &9 &  -5 &  24  \\
2023-04-05 & 22:28:28.8   & 60039.936435 &1.3 & 11.1 & -175  &  -648 &9 &   8 &  13  \\
2023-04-05 & 23:29:09.9   & 60039.978576 &1.2 & 11.1 & -138  &  -660 &9 &  -6 &  13  \\
2023-04-05 & 23:55:50.3   & 60039.997106 &1.2 & 11.1 & -105  &  -666 &13 &   5 &  12  \\
2023-04-06 & 00:50:41.7   & 60040.035197 &1.3 & 11.1 &  -51  &  -683 &13 &  13 &   1  \\
2023-04-06 & 01:17:28.7   & 60040.053796 &1.3 & 11.1 &  -44  &  -697 &9 &  -3 & -11  \\
2023-04-06 & 02:11:47.0   & 60040.091516 &1.4 & 11.1 &  -13  &  -683 &9 & -18 &   5  \\
2023-04-06 & 23:41:11.2   & 60040.986933 &1.2 & 11.1 &  675  &    -3 &13 & -32 &  14  \\
2023-04-07 & 00:46:02.8   & 60041.031968 &1.3 & 11.1 &  698  &     6 &9 &  -7 & -31  \\
2023-04-07 & 01:46:46.1   & 60041.074144 &1.4 & 11.1 &  687  &   105 &9 & -13 &  18  \\
2023-04-07 & 23:09:48.9   & 60041.965139 &1.2 & 11.1 & -124  &   670 &9 & -37 &  -8  \\
2023-04-08 & 00:12:27.2   & 60042.008646 &1.2 & 11.1 & -133  &   687 &13&   7 &  16  \\
2023-04-08 & 00:39:17.6   & 60042.027280 &1.3 & 11.1 & -179  &   647 &9 & -17 & -19  \\
2023-04-08 & 01:37:08.5   & 60042.067454 &1.4 & 11.1 & -227  &   622 &9 & -17 & -32  \\
2023-04-08 & 02:03:37.6   & 60042.085845 &1.4 & 11.1 & -265  &   621 &9 & -34 & -26  \\
2023-04-08 & 20:24:24.1   & 60042.850278 &1.7 & 11.1 & -705  &   -84 &9 &  -9 & -14  \\
2023-04-08 & 21:13:00.1   & 60042.884028 &1.5 & 11.1 & -703  &  -105 &9 & -13 &   4  \\
2023-04-08 & 21:39:51.2   & 60042.902674 &1.4 & 11.1 & -684  &  -131 &9 &   2 &   1  \\
2023-04-08 & 22:27:20.9   & 60042.935648 &1.3 & 11.1 & -670  &  -186 &9 &   6 & -16  \\
2023-04-08 & 22:54:22.3   & 60042.954421 &1.2 & 11.1 & -667  &  -206 &9 &   4 & -14  \\
2023-04-08 & 23:42:23.0   & 60042.987766 &1.2 & 11.1 & -653  &  -227 &11 &  11 & -16  \\
2023-04-09 & 00:09:02.0   & 60043.006273 &1.2 & 11.1 & -636  &  -262 &7 &   8 &  -2  \\
2023-04-09 & 00:56:35.9   & 60043.039294 &1.3 & 11.1 & -636  &  -304 &9 &  -3 & -16  \\
2023-04-09 & 01:23:16.8   & 60043.057824 &1.3 & 11.1 & -615  &  -319 &9 &   8 & -12  \\
2023-04-09 & 02:10:03.1   & 60043.090312 &1.4 & 11.1 & -600  &  -362 &9 &   3 & -20  \\
2023-04-09 & 02:36:32.6   & 60043.108704 &1.6 & 11.1 & -577  &  -352 &9 &  14 &   9   \\
2023-04-09 & 19:53:33.9   & 60043.828854 &1.9 & 11.1 &  193  &  -652 &9 &  17 &  15   \\
2023-04-09 & 20:41:45.4   & 60043.862326 &1.6 & 11.1 &  222  &  -641 &15&   7 &  15  \\
2023-04-09 & 21:08:24.1   & 60043.880833 &1.5 & 11.1 &  251  &  -631 &9 &  15 &  18  \\
2023-04-09 & 21:56:22.8   & 60043.914144 &1.3 & 11.1 &  286  &  -623 &9 &  12 &  13  \\
2023-04-09 & 22:24:01.1   & 60043.933345 &1.3 & 11.1 &  308  &  -624 &9 &  12 &   3  \\
2023-04-09 & 23:11:25.8   & 60043.966262 &1.2 & 11.1 &  350  &  -604 &9 &  19 &   5  \\
2023-04-09 & 23:38:13.6   & 60043.984873 &1.2 & 11.1 &  372  &  -602 &9 &  27 &   0  \\
2023-04-10 & 00:32:02.1   & 60044.022245 &1.2 & 11.1 &  409  &  -578 &9 &  18 &  -2  \\
2023-04-10 & 00:58:49.3   & 60044.040845 &1.3 & 11.1 &  399  &  -558 &5 & -10 &   5  \\
2023-04-10 & 01:46:13.0   & 60044.073762 &1.4 & 11.1 &  449  &  -534 &4 &   8 &   6  \\
2023-04-10 & 02:12:47.7   & 60044.092211 &1.5 & 11.1 &  438  &  -542 &9 & -16 & -12  \\
2023-04-12 & 01:22:18.0   & 60046.057153 &1.4 & 11.1 & -593  &   335 &13&  -7 & -38  \\
2023-04-12 & 02:09:26.9   & 60046.089884 &1.5 & 11.1 & -595  &   331 &9 &  12 &  -9  \\
2023-04-12 & 02:35:57.2   & 60046.108299 &1.9 & 11.1 & -607  &   310 &9 &  20 &   6  \\
2023-04-16 & 00:30:30.4   & 60050.021181 &1.3 & 11.2 & -687  &   -70 &11&   6 & -23  \\
2023-04-16 & 00:57:09.3   & 60050.039688 &1.4 & 11.2 & -684  &   -84 &9&   7 & -15  \\
2023-04-16 & 01:44:28.4   & 60050.072546 &1.6 & 11.2 & -676  &  -104 &5&   9 &   3  \\
2023-04-16 & 02:10:59.5   & 60050.090961 &1.6 & 11.2 & -669  &  -150 &9 &  12 & -21  \\
2023-04-16 & 22:55:42.9   & 60050.955347 &1.2 & 11.2 &   98  &  -658 &9&   2 &  16  \\
2023-04-16 & 23:28:44.7   & 60050.978287 &1.2 & 11.2 &  130  &  -660 &4&   6 &   9  \\
2023-04-16 & 23:48:01.5   & 60050.991678 &1.2 & 11.2 &  147  &  -653 &7&   8 &  14  \\
2023-04-17 & 00:48:48.5   & 60051.033889 &1.4 & 11.2 &  193  &  -655 &4&   4 &   0  \\
2023-04-17 & 01:03:58.3   & 60051.044421 &1.4 & 11.2 &  216  &  -649 &5&  14 &   3  \\
2023-04-17 & 02:15:01.2   & 60051.093762 &1.7 & 11.2 &  260  &  -616 &7&   2 &  16  \\
2023-04-17 & 02:41:11.1   & 60051.111933 &1.9 & 11.2 &  299  &  -639 &9&  21 & -15  \\
2023-04-17 & 19:46:22.1   & 60051.823866 &1.7 & 11.2 &  687  &    54 &4&  -9 &   7  \\
2023-04-17 & 20:36:53.5   & 60051.858947 &1.4 & 11.2 &  670  &    91 &7& -22 &   4  \\
2023-04-17 & 21:01:01.0   & 60051.875706 &1.4 & 11.2 &  667  &   114 &9& -22 &   7  \\
2023-04-17 & 21:38:14.3   & 60051.901551 &1.3 & 11.2 &  670  &   134 &7& -12 &  -4  \\
2023-04-17 & 21:56:30.7   & 60051.914236 &1.2 & 11.2 &  671  &   151 &9&  -8 &  -1  \\
2023-04-17 & 22:35:33.8   & 60051.941354 &1.2 & 11.2 &  663  &   192 &9&  -8 &  10  \\
2023-04-17 & 22:54:14.4   & 60051.954329 &1.2 & 11.2 &  660  &   204 &4&  -6 &   7  \\
2023-04-18 & 00:30:47.1   & 60052.021377 &1.3 & 11.2 &  624  &   280 &9& -14 &   8  \\
2023-04-18 & 00:47:56.1   & 60052.033287 &1.3 & 11.2 &  611  &   289 &13& -21 &   5  \\
2023-04-18 & 01:26:17.0   & 60052.059919 &1.5 & 11.2 &  602  &   325 &9& -15 &  13  \\
2023-04-18 & 01:46:14.4   & 60052.073773 &1.6 & 11.2 &  593  &   340 &11& -16 &  13  \\
2023-04-18 & 02:23:12.3   & 60052.099444 &1.8 & 11.2 &  578  &   366 &9 & -15 &  14  \\
2023-04-18 & 19:35:14.5   & 60052.816134 &1.7 & 11.2 & -178  &   644 &13& -18 & -10  \\
2023-04-18 & 20:17:18.4   & 60052.845347 &1.5 & 11.2 & -213  &   649 &13 & -19 &   4  \\
2023-04-18 & 20:36:48.5   & 60052.858889 &1.4 & 11.2 & -219  &   631 &13& -14 & -11  \\
2023-04-18 & 21:13:47.5   & 60052.884572 &1.3 & 11.2 & -251  &   622 &4 & -12 &  -8  \\
2023-04-18 & 21:33:53.6   & 60052.898530 &1.3 & 11.2 & -279  &   610 &13 & -23 & -15  \\
2023-04-18 & 22:16:10.1   & 60052.927894 &1.2 & 11.2 & -301  &   598 &13 & -17 & -15  \\
2023-04-18 & 22:36:03.6   & 60052.941701 &1.2 & 11.2 & -318  &   594 &9& -20 & -12  \\
2023-04-18 & 23:13:58.5   & 60052.968032 &1.2 & 11.2 & -338  &   574 &4 &  -6 & -15  \\
2023-04-18 & 23:34:19.3   & 60052.982164 &1.2 & 11.2 & -359  &   567 &4 & -12 & -14  \\
2023-04-19 & 00:14:57.4   & 60053.010382 &1.3 & 11.2 & -392  &   559 &7& -16 &  -5  \\
2023-04-19 & 00:36:09.0   & 60053.025104 &1.3 & 11.2 & -403  &   548 &5& -12 &  -6  \\
2023-04-19 & 01:12:49.5   & 60053.050567 &1.4 & 11.2 & -427  &   528 &9& -11 &  -9  \\
2023-04-19 & 01:30:52.1   & 60053.063102 &1.5 & 11.2 & -440  &   529 &13& -16 &  -2  \\
2023-04-19 & 02:09:21.2   & 60053.089826 &1.7 & 11.2 & -459  &   492 &9&  -6 & -16  \\
2023-04-19 & 02:23:58.2   & 60053.099977 &1.9 & 11.2 & -481  &   491 &9& -19 &  -9  \\
\end{longtable}                                     
\twocolumn}                                         
                                                    

\end{document}